\documentclass[aps,pra,twocolumn,showpacs]{revtex4}

\usepackage{amsmath}
\usepackage{bm}
\usepackage{graphicx}

\begin{document}

\title{Low-temperature quenching of one-dimensional localised Frenkel
excitons}

\author{A.\ V.\ Malyshev}
\thanks{On leave from Ioffe Physiko-Technical Institute,
26 Politechnicheskaya str., 194021 Saint-Petersburg, Russia}
\affiliation{GISC, Departamento de F\'{\i}sica de Materiales, Universidad
Complutense, E-28040 Madrid, Spain}

\author{V.\ A.\ Malyshev}
\thanks{On leave from ``S.I. Vavilov State Optical Institute'',
Saint-Petersburg, Russia}
\affiliation{GISC, Departamento de F\'{\i}sica de Materiales, Universidad
Complutense, E-28040 Madrid, Spain}

\author{F.\ Dom\'{\i}nguez-Adame}
\affiliation{GISC, Departamento de F\'{\i}sica de Materiales, Universidad
Complutense, E-28040 Madrid, Spain}

\date{\today}


\begin{abstract}

We present a theoretical analysis of low-temperature quenching of
one-dimensional Frenkel excitons that are localised by moderate on-site
(diagonal) uncorrelated disorder. Exciton diffusion is considered as an
incoherent hopping over localization segments and is probed by the exciton
fluorescence quenching at point traps. The rate equation is used to
calculate the temperature dependence of the exciton quenching. The
activation temperature of the diffusion is found to be of the order of the
width of the exciton absorption band. We demonstrate that the intra-segment
scattering is extremely important for the exciton diffusion. We discuss also
experimental data on the fast exciton-exciton annihilation in linear
molecular aggregates at low temperatures.

\end{abstract}


\maketitle

\section{Introduction}
\label{Intro}

Since the seminal works by Jelley~\cite{Jelley36} and Scheibe~\cite{Scheibe36},
the concept of Frenkel
excitons~\cite{Frenkel31,Davydov71} has been admitted for the explanation of the
remarkable optical properties of molecular aggregates. The most surprising
among them are the appearance of a narrow and intense line in the red-wing
of the absorption spectra (J-band) and the increase of the oscillator
strength of the optical transition at cryogenic temperatures by almost two
orders of magnitude~\cite{deBoer90,Fidder90,Fidder91}. During the nineties,
a significant progress in understanding of optical dynamics in J-aggregates
was achieved (for details see the recent
reviews~\onlinecite{Spano94,AdvMat95,Kobayashi96} and references therein).

In the recent paper~\onlinecite{Scheblykin00}, the anomalously fast low-temperature 
mobility of excitons in linear molecular aggregates was reported. 
The authors of Ref.~\onlinecite{Scheblykin00} studied the exciton-exciton
annihilation in aggregates of the triethylthiacarbocyanine salt of
3,3'-bis(sulfopropyl)-5,5'-dichloro-9-ethylthiacarbocyanine (THIATS). The
process of annihilation was found to activate at low temperatures, $T < 20$
K (which is small compared to the width of the J-band, 82 cm$^{-1}$) and,
what is more important, at extremely low concentrations of excitons. This
finding implies that excitons can move at low temperatures over large
distances.

A great deal of theoretical work has also been devoted to description of the
optical properties of J-aggregates~\cite{Kobayashi96}. Surprisingly, the simplest
one-dimensional (1D) tight-binding model with on-site (diagonal) disorder
provides a good basis for understanding complex optical dynamics in these
systems. The goal of the present paper is to study whether this model
describes the experimentally observed high mobility of 1D Frenkel excitons in
J-aggregates at low-temperatures. To the best of our knowledge, this problem
has not been discussed in the literature yet. We use the effect of quenching
of the exciton fluorescence to probe the exciton mobility. The term "low
temperature" refers to the temperatures lower or of the order of the
magnitude of the J-band width. Higher temperatures are beyond the scope of
this paper.

The paper is organized as follows. In Sec.~\ref{Model}, we present our
microscopic model. In Sec.~\ref{Qualit}, the low-temperature exciton
diffusion over manifolds of the localised states and quenching are discussed
qualitatively. The results of numerical simulations of the exciton
fluorescence quenching, obtained on the basis of the rate equation approach,
are the contents of Sec.~\ref{Numerics}. In Sec.~\ref{Exp-data}, we discuss
the experimental data on the exciton-exciton annihilation in THIATS aggregates.
Section~\ref{Concl} concludes the paper with a summary of the main results.

\section{Description of the model}
\label{Model}

We model a J-aggregate by $N$ ($N \gg 1$) optically active two-level
molecules forming a regular in space 1D open chain. The corresponding
Frenkel exciton Hamiltonian reads~\cite{Davydov71} (for the sake of
simplicity only the nearest-neighbor interaction is considered)
\begin{equation}
    H = \sum_{n=1}^{N}\> E_n |n\rangle \langle n|
    - J \sum_{n=1}^{N - 1}\> (|n + 1\rangle \langle n|
    + |n\rangle \langle n + 1| )\ .
    \label{H}
\end{equation}
Here $E_n$ is the excitation energy of the $n$-th molecule, $|n \rangle$
denotes the state vector of the $n$-th excited molecule. The energies $E_n$
are assumed to be Gaussian uncorrelated (for different sites) stochastic
variables distributed around the mean value $\omega_0$ (which is set to zero
without loosing generality) with the standard deviation $\Delta$. The
hopping integral, $-J$, is considered to be non-random and negative ($J
> 0$), which corresponds to the case of J-aggregates (see, e.g.,
Ref.~\cite{deBoer90}). In this case the states coupled to the light are
those close to the bottom of the exciton band. We consider moderate disorder
($\Delta < J$) in what follows. This implies that the exciton eigenstates
$\varphi_{\nu} \, (\nu = 1, 2, \ldots , N)$ found from
\begin{equation}
\sum_{n=1}^N H_{nm}\varphi_{\nu m} = \varepsilon_\nu \varphi_{\nu n} \ ,
\qquad
H_{nm}=\langle n|H|m \rangle
\label{k}
\end{equation}
are extended over relatively large segments of the chain. However, the
typical size of these localisation segments, $N^*$, is small compared to the
chain length $N$.

Having been excited into an eigenstate $\nu$, an exciton cannot hop to other
eigenstates if coupling to vibrations is not taken into account. This
coupling causes the {\it incoherent} hopping of excitons from one
eigenstate to another. We take the hopping rate from the state $\nu$ to the
state $\mu$ in the following form (see, e.g., Ref.~\onlinecite{Leegwater97})
\begin{multline}
W_{\mu\nu} = W_0\ S(\left|\varepsilon_\nu - \varepsilon_{\mu}\right|)
\,\sum_{n=1}^N \varphi_{\nu n}^2 \varphi_{n\mu}^2 \ \times\\ \times
\left\{
\begin{array}{lr}
n(\varepsilon_\mu - \varepsilon_{\nu}), &\quad \varepsilon_\mu > \varepsilon_{\nu}\\
1+n(\varepsilon_{\nu} - \varepsilon_\mu), &\quad \varepsilon_\mu < \varepsilon_{\nu}
\end{array}
\right. \ .
\label{1Wkk'}
\end{multline}
Here, the constant $W_0$ characterizes the amplitude of the hopping and
$n(\varepsilon) = [\exp(\varepsilon/T) - 1]^{-1}$ is the occupation number of
the vibration mode with the energy $\varepsilon$ (the Boltzmann constant is
set to unity). Due to the presence of the $n(\varepsilon)$ and
$1+n(\varepsilon)$ factors, the rate $W_{\mu\nu}$ meets the principle of
detailed balance: $W_{\mu\nu} = W_{\nu\mu}\exp[(\varepsilon_{\nu} -
\varepsilon_\mu)/T]$. Thus, in the absence of decay channels, the eventual
exciton distribution is the Boltzmann equilibrium distribution. The sum over
sites in (\ref{1Wkk'}) represents the overlap integral of exciton
probabilities for the states $\mu$ and $\nu$. The spectral factor
$S(|\varepsilon_\nu - \varepsilon_{\mu}|)$ depends strongly on the
particular details of the exciton-phonon coupling as well as on the density
of states of the medium into which the aggregate is embedded. The study of
these details is beyond the scope of the present paper. We use the linear
form $S(|\varepsilon_\nu -\varepsilon_{\mu}|)=|\varepsilon_\nu -
\varepsilon_{\mu}|/J$ of this factor, which accounts for the reduction of the
hopping in the long-wave acoustic limit~\cite{Bednarz01}. The hopping rate
in the form of Eq.~(\ref{1Wkk'}) was used in
Ref.~\cite{Shimizu01,Bednarz01} to successfully describe the 1D
exciton thermalization.

The mobility of excitons can be probed by the quenching of the exciton
fluorescence, which is due to point traps. The quenching rate of the exciton
state $\nu$ is assumed to be proportional to the probability of finding the
exciton at trap sites
\begin{equation}
\Gamma_\nu = \Gamma \sum_{i=1}^{N_q} |\varphi_{\nu i}|^2 \ ,
\label{Gamma}
\end{equation}
where $\Gamma$ is the amplitude of exciton quenching and the sum runs over
positions of the $N_q$ traps. We assume also that point traps do not change
neither the disorder configuration nor the exciton eigenfunctions
$\varphi_{\nu i}$.

We describe the process of the exciton trapping by means of the rate
equation:
\begin{equation}
{\dot P}_\nu = -(\gamma_\nu + \Gamma_\nu) P_\nu
+ \sum_{\mu =1}^N (W_{\nu\mu}\,P_{\mu}
- W_{\mu\nu}\,P_\nu)\ ,
\label{P_nu}
\end{equation}
where $P_\nu$ is the population of the $\nu$th exciton eigenstate
and the dot denotes the time derivative, $\gamma_\nu = \gamma
(\sum_{n=1}^N \varphi_{\nu n})^2$ is the spontaneous emission rate of
the $\nu$th exciton state, while $\gamma$ is that of a monomer. The initial
total population is normalized to unity: $\sum_\nu P_\nu (0) = 1$.

The temperature dependence of the exciton quenching is calculated as
follows. We admit the definition of the exciton fluorescence
decay time $\tau$ as the total population integrated over time (see, for
instance, Ref.~\cite{Bednarz02}). In the presence of disorder, it
has to be averaged over disorder configurations and trap
positions. Then, the equation for $\tau$ reads
\begin{multline}
\tau = \int_0^\infty dt \int dE \, \left\langle \sum_{\nu = 1}^N
\delta(E - E_\nu)P_\nu(t) \right\rangle = \\
= \int_0^\infty dt \, \left\langle \sum_{\nu = 1}^N P_\nu(t) \right\rangle\ ,
\label{tau}
\end{multline}
where angle brackets denote averaging. The decay time is calculated for
aggregates with traps (denote it as $\tau$) and without traps (denote
it as $\tau_0$) for the same set of disorder configurations. Recall that,
according our assumption, traps do not change the exciton eigenfunctions.
The quenching rate is then defined as
\begin{equation}
W_q = \frac{1}{\tau} -\frac{1}{\tau_0} \ .
\label{Wq}
\end{equation}
This quantity carries information about the diffusion rate and is the object
of our analysis.

The definition of the decay rate as the integrated total population allows
for considerable simplification of the calculation procedure. We write the
solution of Eq.~(\ref{P_nu}) in a formal matrix form
\begin{equation}
P_\nu = \sum_{\mu=1}^N \left( e^{-{\hat R}t} \right)_{\nu\mu}P_{\mu}(0) \ ,
\label{FormSol}
\end{equation}
where
\begin{equation}
R_ {\nu\mu} =
\left ( \gamma_\nu + \Gamma_\nu + \sum_{\mu=1}^N W_{\mu\nu}
\right ) \delta_{\mu\nu} - W_{\nu\mu} \ .
\label{R}
\end{equation}
After the substitution of (\ref{FormSol}) into Eq.~(\ref{tau}) and
integration over time, $\tau$ can be expressed in terms of the
$\hat{R}$-matrix:
\begin{equation}
\tau =  \left\langle \, \sum_{\nu,\mu=1}^N  \left( {\hat R}^{-1}
\right )_{\nu\mu} P_\mu(0) \right\rangle \ .
\label{1tau}
\end{equation}
Clearly, obtaining $W_q$ only requires the calculation of the inverse matrix
${\hat R}^{-1}$ for each realization of disorder rather than the tedious
calculation of the kinetics. The inverse matrix is to be found twice: for an
aggregate with and without traps. Note that the decay time in the absence of
traps, $\tau_0$, also depends on temperature (see, for example,
Ref.~\cite{Bednarz01}).

\section{Qualitative picture}
\label{Qualit}

\begin{figure}[ht]
\includegraphics[width=\columnwidth,clip]{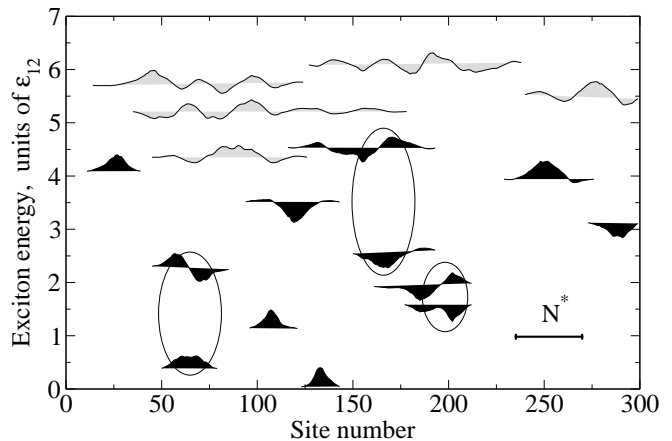}
\caption{The energy structure of the exciton levels in the vicinity of
the bottom of the exciton band. The states are obtained by diagonalization
of the Hamiltonian~(\protect\ref{H}) for a linear chain of 300 sites and the
disorder magnitude $\Delta= 0.1J$. The baseline of each state represents its
energy in units of $\varepsilon_{21}$. The origin of the exciton energy is
set to the lowest energy for the realization. The wave functions are in
arbitrary units. It is clearly seen that the lower states can be grouped
into local manifolds. The states within each such manifold are localised at
the same segment; they overlap well with each other and overlap much weaker
with the states of other manifolds.}
\label{fig1}
\end{figure}

At low temperatures, excitons occupy states at the bottom of the exciton
band. Therefore, this part of the exciton energy spectrum determines the
low-temperature exciton transport. Below, we recall briefly the concept of
local (hidden) energy structure of localised 1D
excitons~\cite{Malyshev91,Malyshev95,Malyshev99a}, which was proved to exist
in the vicinity of the band bottom~\cite{Malyshev01a,Malyshev01b}. According
to this concept, the low-energy one-exciton eigenfunctions obtained for a
fixed realization of disorder can be grouped into local manifolds of two (or
sometimes more) states that are localised at the same chain segment of
typical length $N^*$ (in units of the lattice constant) which scales with
disorder as follows~\cite{Malyshev01a}
\begin{equation}
N^* = 8.71\,\left(\frac{\Delta}{J}\right)^{-0.67}\ .
\label{scalingA}
\end{equation}
It turns out that the structure of the exciton states in each local manifold is
very similar to the structure of the lower states of a regular (non-disordered)
chain of length $N^*$. In particular, the lowest state in a manifold has a wave
function without nodes within its localisation segment. Such a state can be
interpreted as the {\it local} ground state of the segment. It carries large
oscillator strength, approximately $N^*$ times larger than that of a
monomer, so that the typical spontaneous emission rate is $\gamma^* =
\gamma N^*$. The second state in the manifold has a node within the
localisation segment and looks like the first {\it local} excited state of
the segment (see the states filled with black color in Fig.~\ref{fig1}). Its
oscillator strength is typically an order of magnitude smaller than that of
the local ground state. It is important to note that, contrary to the
eigenstates from the same manifold, the states from different manifolds
overlap weakly. The energies of local ground states are distributed within
the interval $\sigma_{11}$ that is larger than the typical energy spacing
$\varepsilon_{12}$ between the levels in a local
manifold~\cite{Malyshev01a}:
\begin{subequations}
\begin{equation}
\sigma_{11} = 0.67\,J\,\left(\frac{\Delta}{J}\right)^{1.33}\ ,
\label{scalingC}
\end{equation}
\begin{equation}
\varepsilon_{12} = 0.40\,J\,\left(\frac{\Delta}{J}\right)^{1.36}\ .
\label{scalingD}
\end{equation}
\label{scaling}
\end{subequations}
For this reason, the local energy structure cannot be seen either in the
density of states (DOS) or in the linear
absorption spectra. However, it determines the nonlinear optical response of
the system~\cite{Malyshev95,Minoshima94,Knoester96,Bakalis99}.

Higher states are more extended than the {\it local} states as the
localisation length increases with energy. Therefore, the higher states cannot
be included into any particular local manifold: their wave functions
cover more than one $N^*$-molecule segments (see the states filled with gray
color in Fig.~\ref{fig1}). The typical energy spacing between these higher
states and the covered local states is of the order of $\varepsilon_{12}$
Thus, the energy $\varepsilon_{12}$ is expected to be the activation energy
for the exciton diffusion.

It is clear from the above arguments that the energy structure of the lower
exciton states is of importance for understanding the features of the
low-temperature exciton transport. On the basis of the local energy
structure concept, two types of exciton hopping over the {\it local} states
can be distinguished: intra-segment hopping and inter-segment one, involving
the states of the same local manifold and of different manifolds,
respectively. As the states from different local manifolds overlap weakly
(see Fig.~\ref{fig1}), only inter-segment hops to adjacent segments are of
importance. The disorder scaling of the overlap integrals $I_{\mu\nu} =
\sum_n \varphi_{\mu n}^2 \varphi_{\nu n}^2$ for the local states of the same
and adjacent segments was obtained in Ref.~\onlinecite{Malyshev01b}:
\begin{subequations}
\begin{equation}
I_{12} = 0.14\ \left(\frac{\Delta}{J}\right)^{0.70}
\label{scalingE}
\end{equation}
\begin{equation}
I_{\nu^\prime 1} = I_{\nu^\prime 2} \approx 0.0025\
\left(\frac{\Delta}{J}\right)^{0.75}  \ .
\label{scalingF}
\end{equation}
\label{scalingI}
\end{subequations}

\noindent
Hereafter, the indices 1 and 2 label the local states of the same segment
while those with primes label the local states of a different adjacent
segment. As follows from Eqs.~(\ref{scalingE})-(\ref{scalingF}), the
intra-segment overlap integral is typically more than an order of magnitude
larger than the inter-segment one. The intra-segment hops do not result in
the spatial displacement of excitons; they correspond to the intra-segment
relaxation. Only the inter-segment hopping gives rise to the spatial motion
of excitons. Nevertheless, we show that both types of hops are important
for understanding the features of the low-temperature exciton transport.

The overlap integrals between the local states of a segment and the higher
states which are extended over this segment are of the order of $I_{12}$.
This fact implies that even at $T < \varepsilon_{12}$, the hops via these
higher states can be more efficient than the inter-segment hops over the
local states.

\subsection{Zero temperature}

\begin{figure}[ht]
\includegraphics[height=\columnwidth,clip,angle=90]{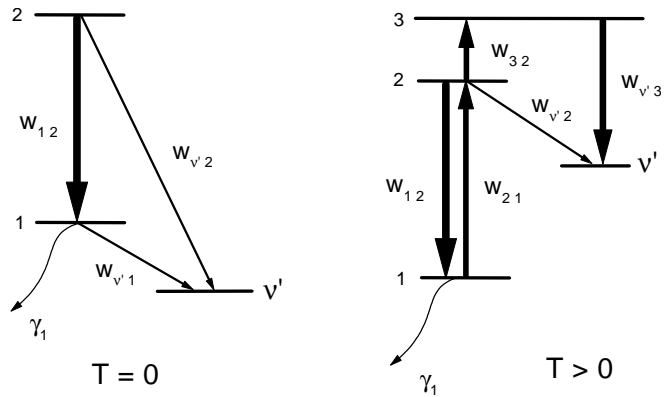}
\caption{Schematic view of exciton hoppings at zero and non-zero
temperatures. The indices $1$ and $2$ label the {\it local} ground and the
first {\it local} excited states of the same segment. The $\nu^\prime$ state
is localised at an adjacent segment. The index 3 labels a higher state which
overlaps well with the states of two adjacent segments. Hops are shown by the
straight arrows; the arrow thickness represents the hopping rate magnitude.
The spontaneous emission rate $\gamma_1$ is depicted by thin wavy arrows.}
\label{fig2}
\end{figure}

At zero temperature an exciton can hop only down to lower states. Let us
assume that it is in the {\it local} excited state 2.  Then it can only hop
to the {\it local} ground state of the same segment 1 or to a lower state
$\nu^\prime$ localised at an adjacent segment (see Fig.~\ref{fig2}, $T=0$).
Because the intra-segment hopping is faster than the inter-segment one,
first, the exciton hops down to the local ground state $1$ with the typical
energy loss $\varepsilon_{12}$ ($\varepsilon_{12}$ being the mean energy
spacing in the local discrete energy structure, see Fig.~\ref{fig2}, $T=0$).
From the local ground state, the exciton can hop only to a state
$\nu^\prime$ of an adjacent segment provided that
$\varepsilon_{\nu^\prime} < \varepsilon_{1}$ and the spontaneous emission
rate of the local ground state $\gamma_1$ is small compared to the
intra-segment hopping rate $W_{\nu^\prime 1}|_{T=0}$. Hereafter, such a
relationship between these rates is referred to as the limit of fast
diffusion; only this limit is considered in this work. The typical energy
loss during such sideways hop is of the order of the width of the local
ground states distribution, $\sigma_{11}$ ($\sigma_{11}$ is about the J-band
width). Thus, already after one such sideways hop the exciton resides in a
state in the tail of the DOS. Therefore, the number of states with even
lower energies decreases dramatically, which results in a strong increase in
the typical distance to those states and in a suppression of the
probability to hop further sideways. Then the exciton either relaxes to a
lower state of the same segment (if there is one) or decays spontaneously,
i.e., this type of the spatio-energetic diffusion (towards lower energies)
stops very quickly. Note that this diffusion would manifest itself in the
red shift of the exciton emission spectrum relative to the absorption
spectrum. The experimental data show that such red shift is either
absent~\cite{deBoer90,Fidder90} or is smaller than the J-band
width.~\cite{Scheblykin00,Kamalov96}. These experimental findings
unambiguously indicate that at low temperatures, $T \ll \sigma_{11}$,
excitons make few hops before they decay due to the spontaneous emission, as
was argued in Refs.~\onlinecite{Malyshev99b,Malyshev00,Ryzhov01}.
Consequently, the zero-temperature exciton quenching is expected to be weak
provided the concentration of quenchers is low, the case we are interested
in.

\subsection{Non-zero temperature}

At non-zero temperatures ($0 < T \lesssim \varepsilon_{12}$), an exciton can
also hop up in energy. Consider an exciton in one of the lower states in the
tail of the DOS, e.g. in the {\it local} ground state $1$ (see
Fig.~\ref{fig2}, $T > 0$). For the reasons discussed above, first, the
exciton hops up to the first {\it local} excited state $2$ of the same
segment, provided the hopping rate for the considered temperature is larger
than the spontaneous emission rate $\gamma_1$ of the initial state 1. During
this process the exciton typically gains the energy $\varepsilon_{12}$. As
$\varepsilon_{12}$ is of the order of $\sigma_{11}$~\cite{Malyshev01a},
already after the first hop up the exciton leaves the tail of the DOS and,
hence, it is likely to have a lower state $\nu^\prime$ localised at an
adjacent segment. A hop down to this state with loss in energy is favorable
and results in the spatial displacement of the exciton, i.e., in the exciton
diffusion. We stress that although only sideways hops result in the spatial
displacement of the exciton, it is the initial hop up from the {\it local}
ground state $1$ to the {\it local} excited state $2$ that triggers the
diffusion.

Another way for the exciton to hop sideways to the state $\nu^\prime$ is via
the higher state $3$ that overlaps well with both states $2$ and $\nu^\prime$
(see Fig.~\ref{fig2}, $T > 0$). As it has been mentioned, such hops compete
with the sideways hops over the local states; although the hop up to the state
$3$ is thermally unfavorably, the overlap integral for this hop, $I_{31}$,
is large compared to that for an inter-segment hop, $I_{\nu^\prime 1}$.
We show later that this channel of diffusion becomes efficient indeed
even at relatively low temperature.

\section{Numerical results and discussion}
\label{Numerics}

In this section, we discuss the results of numerical calculation of the
quenching rate $W_q$. In this paper, we consider the initial condition where
the leftmost local ground state is excited while a single trap is located in
the center of the localisation segment of the rightmost local ground state.
In this case, the exciton quenching is most affected by diffusion, as the
created exciton has to travel over almost the whole chain to be quenched.
Thus, the exciton quenching at low concentration of traps can be studied.
The quenching rate was calculated as described in section \ref{Model} for
the parameter set corresponding to the limit of fast diffusion and effective
quenching (the later limit is defined below). In all calculations we set
$W_0=1$ and choose the parameter $\gamma$ so that the typical inter-segment
down-hopping rate $W_{1^\prime 1}|_{T=0} \sim
W_0\,(\sigma_{11}/J)\,I_{\nu^\prime 1}$ is large compared to the typical
spontaneous emission rate of a local ground state $\gamma^*$ (the limit of
fast diffusion). $\Gamma$ is chosen so that the typical quenching rate
$\Gamma^*$ is greater than the typical intra-segment down-hopping rate
$W_{12}|_{T=0} \sim W_0\ (\varepsilon_{12}/J)\ I_{12}$ (the limit of
effective quenching). This ensures that once an exciton hops to a local
state of the segment with the trap, it is quenched almost instantaneously.
More specifically, for each magnitude of the disorder $W_{1^\prime
1}|_{T=0} = 10\,\gamma^*$ and $\Gamma^* = 10\,W_{12}|_{T=0}$ : 
$\gamma=5\times 10^{-8}$ and $\Gamma=0.2$ for $\Delta=0.1\,J$;  
$\gamma=4\times 10^{-7}$ and $\Gamma=0.5$ for $\Delta=0.2\,J$;  
$\gamma=1\times 10^{-6}$ and $\Gamma=0.9$ for $\Delta=0.3\,J$. 
Calculations were performed for $N=1000$ and 100 realizations of the disorder.

\begin{figure}[ht]
\includegraphics[width=\columnwidth,clip]{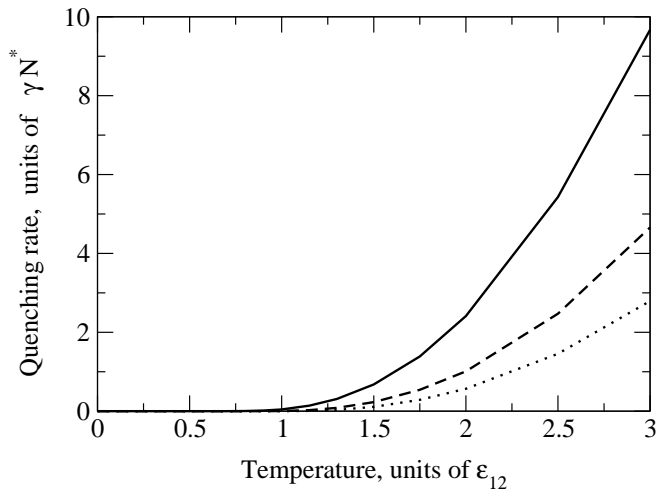}
\caption{Temperature dependence of the quenching rate $W_q$ calculated
for a linear chain length $N = 1000$ and different magnitudes of the
disorder $\Delta$: solid line --- $\Delta=0.1\,J$, dashed line ---
$\Delta=0.2\,J$, dotted line --- $\Delta=0.3\,J$. 
The rate equation parameters were chosen so that for each magnitude of
disorder $W_{1^\prime 1}|_{T=0} = 10\,\gamma^*$ and $\Gamma^* =
10\,W_{12}|_{T=0}$ :
$\gamma=5\times 10^{-8}$ and $\Gamma=0.2$ for $\Delta=0.1\,J$; $\gamma=4\times
10^{-7}$ and $\Gamma=0.5$ for $\Delta=0.2\,J$; $\gamma=1\times 10^{-6}$ and
$\Gamma=0.9$ for $\Delta=0.3\,J$.
The averaging is performed over 100 disorder realizations.
For each realization of the disorder, the leftmost local ground state is
excited, while the only trap is located in the center of the localisation
segment of the rightmost {\it local} ground state.}
\label{fig3}
\end{figure}

Figure \ref{fig3} shows the temperature dependence of the quenching rate
$W_q$ for the parameter sets specified above. In the plot, the quenching
rate is given in units of the typical exciton radiative rate
$\gamma^* = \gamma N^*$. The temperature is given in units of the mean
energy spacing in the local energy structure $\varepsilon_{12}$. Note that
both $N^*$ and $\varepsilon_{12}$ depend on $\Delta$ as described
by~(\ref{scalingA}) and~(\ref{scalingD}). Figure \ref{fig3} demonstrates
very clearly that for all considered values of $\Delta$ at temperatures
lower than $\varepsilon_{12}$ the quenching rate is vanishing. This
indicates that at these temperatures the exciton cannot reach the quencher
during its lifetime and decays due to the spontaneous emission. On the
contrary, just after the temperature exceeds approximately
$\varepsilon_{12}$ the quenching becomes noticeable: the exciton partly
diffuses to the trap where it decays mostly due to quenching. Specifically,
temperature of the order of $2\,\varepsilon_{12}$ are required for the
quenching to become as effective as the spontaneous emission: $W_q \sim
\gamma^* = \gamma N^*$.

It is useful to estimate the effective sideways hopping rate $W$, which is
required to reach the quenching level $W_q \sim \gamma^*$. To do this, one
can consider the sequence of localization segments as an effective chain of
"sites", the typical number of which is equal to the number of segments,
$N_s= N/N^*$; the mean spacing between these "sites" is $N^*$. The exciton
diffusion coefficient is then estimated as $D \sim W{N^*}^2$ (the lattice
constant is set to unity). For the quenching to be as effective as the
spontaneous decay, the exciton has to reach the quencher (located on the
opposite side of the chain) during the lifetime ${\gamma^*}^{-1}$, i.e.,
it has to diffuse over the distance $N_s$ during this time. Equating the
diffusion length $\sqrt{D/\gamma^*}$ to $N$, we obtain the estimate for
the required diffusion rate $W$: $W \sim \gamma^* (N/N^*)^2$. The
localisation length $N^*$ is equal to 38, 25 and 18 for $\Delta$ = 0.1, 0.2
and 0.3, respectively. Thus, the corresponding diffusion rates $W$ are
estimated as $600 \gamma^*, \, 1600\gamma^*$ and $2500 \gamma^*$.
These values are about of two orders of magnitude larger than the rates of
sideways hops over the local states, taken to be $10 \gamma^*$ in all
calculations. This indicates that when the quenching rate becomes comparable
to the spontaneous emission rate, the exciton does not hop between the local
states of adjacent segments (with the typical rate ($W_{\nu^\prime 1} \sim
10\gamma^*$). It rather hops via the higher states that extend over more
than one $N^*$-molecule segments (see the discussion in Sec.~\ref{Qualit}).
The hopping rate via such states for $T \sim 2\varepsilon_{12}$ is of the
order of $W_{12}$ which is about two orders of magnitude larger than
$W_{\nu^\prime 1}$.

\section{Analyzing the fast exciton-exciton annihilation}
\label{Exp-data}

In Ref.~\cite{Scheblykin00} the anomalously fast low-temperature diffusion of
Frenkel excitons in linear aggregates of THIATS molecules was reported. The
unit cell in these aggregates contains two THIATS molecules. Because of this
fact, the absorption spectra of THIATS aggregates reveal two bands, so-called
H-band and J-band~\cite{Scheblykin00a}. The former, intensive and widely
broadened (the width being about 1000cm$^{-1}$), results from the optical
transition from the ground state of the aggregate to the top of the exciton band.
The latter, much less intensive and narrower (the width being 82cm$^{-1}$),
is due to the optical transition from the ground state to the bottom of the
exciton band. Contrary to the H-band, the J-band is visible in exciton
fluorescence spectra.

The authors of Ref.~\cite{Scheblykin00} studied the exciton-exciton annihilation
in THIATS aggregates by measuring the exciton fluorescence decay after
excitation into the centre of the exciton band (the whole width being about
3000cm$^{-1}$). It was found that this effect is pronounced even at $T = 5$K
(3.5 cm$^{-1}$) and at a very low intensity of excitation. In order to explain
the experimental data, the authors assumed that excitons travel over about $10^4$
dye molecules during their lifetime to meet each other and annihilate. They
found also that the activation energy for the exciton diffusion was 15K (10.5
cm$^{-1}$) and considered this energy to be the typical energy difference
between the states of adjacent localisation segments.

The fact that the exciton-exciton annihilation is very sensitive to temperature
indicates that this process starts after the exciton intra-band relaxation to
the states forming the J-band. Therefore, the annihilation process
involves only the J-band states and the difference in band structure between
THIATS aggregates and J-aggregates is probably unimportant. Furthermore, the
exciton-exciton annihilation can be treated similarly to the exciton quenching:
one of the two excitons can be considered as an immobile trap for the other
while the other diffuses twice as fast. Thus, our model is applicable to
analyzing the exciton-exciton annihilation in THIATS aggregates.

As reported in Ref.~\cite{Scheblykin00}, the fluorescence spectrum of THIATS
aggregates is narrowed by approximately 26 cm$^{-1}$ and experiences a red
shift of 23 cm$^{-1}$ as compared to the J-band. These findings indicate
that the excitons make sideways hops during their lifetime, i.e., the rate
of sideways hops over local states is larger than the exciton spontaneous
emission rate. This indicates that the conditions for the exciton diffusion
in THIATS aggregates are similar to those studied in the present paper (the
limit of fast diffusion). On the basis of this analogy, the activation
energy for the exciton diffusion is expected to be of the order of 82
cm$^{-1}$ rather than the reported 10.5 cm$^{-1}$. The typical size of
localisation segment in THIATS aggregates is $N^* =
30$~\cite{Scheblykin00a}. In the model we are dealing with, this corresponds
to the disorder magnitude $\Delta = 0.2 J$. As it follows from our numerical
data, the exciton quenching is vanishingly small for temperatures $T \sim
(10.5/82) \times \varepsilon_{12}$. Thus, the model of the
temperature-activated hopping over localisation segments, proposed in
Ref.~\cite{Scheblykin00} for the explanation of fast exciton diffusion at
low temperature, is questionable. Understanding the observed fast
low-temperature exciton-exciton annihilation still remains an open question.

\section{Summary}
\label{Concl}

We analyzed theoretically the features of the low-temperature diffusion of
1D Frenkel excitons localised by a moderate diagonal disorder. In this case
the low-energy exciton wave functions are extended over relatively large
segments of typical size $N^*$ ($1 \ll N^* \ll N$). We considered the
exciton motion as {\it incoherent} hops over localised states.
The exciton diffusion was probed by the exciton quenching by a trap
that was located at one end of the chain while the exciton is initially
located at the other end of the chain. For this initial condition, the
quenching is most affected by the diffusion as the exciton has to travel
over almost the whole chain to be quenched. Exciton quenching was described
by the rate equation. Numerical simulations confirm our qualitative
finding that exciton diffusion is activated at the temperature that is
approximately equal to the mean spacing in the local discrete energy
structure. This temperature is of the order of the J-band width.
According to the general belief, at such temperatures the exciton
diffuses over $N^*$-molecule localisation segments. We show however that
it diffuses mostly over higher states which extend over few
such segments. The latter provide the natural energy scale and determine
the activation energy of the diffusion. We demonstrate therefore that the
intra-segment scattering is extremely important for the diffusion.


\acknowledgments

This work was supported by the DGI-MCyT (Project MAT2000-0734). A.~V.~M. and
F.~D.~A. acknowledge support from CAM (Project 07N/0075/2001). V.~A.~M.
acknowledges support from MECyD (Project SAB2000-0103) as well as through a
NATO Fellowship.


\end{document}